\documentclass[12pt]{article}
\usepackage[russian]{babel}
\usepackage{graphicx}
\newcommand{\be}{\begin{equation}}
\newcommand{\ee}{\end{equation}}
\newcommand{\ba}{\begin{eqnarray}}
\newcommand{\ea}{\end{eqnarray}}

\begin{document}
\begin{center}
{\bf {Nambu-Poisson dynamics with some applications
}}
\end{center}
\begin{center}
 Nugzar Makhaldiani
\end{center}
\begin{center}
Joint Institute for Nuclear Research\\
 Dubna, Moscow Region, Russia\\
e-mail address:~~mnv@jinr.ru
\end{center}
\centerline{\bf Abstract}
Short introduction in NPD with several applications to (in)finit dimensional problems of mechanics, hydrodynamics,
M-theory and quanputing is given.


e
%

\begin{flushright}
 \vspace{.5cm}
Nabu --
Babylonian God\\ of
Wisdom and Writing.
\end{flushright}


The Hamiltonian mechanics (HM) is in the fundamentals of
mathematical description of the physical theories \cite{fad}. But
HM is in a sense blind; e.g., it does not make a difference
between two opposites: the ergodic Hamiltonian systems (with just
one integral of motion) \cite{ergo} and (super)integrable
Hamiltonian systems (with maximal number of the integrals of
motion).

Nabu mechanics (NM) \cite{Nambu,Whittaker} is a
proper generalization of the HM, which makes the difference
between dynamical systems with different numbers of integrals of
motion explicit (see, e.g.\cite{NPD} ).\newline

\section{ Hamiltonization of
dynamical systems}

Let us consider a general dynamical system described by the
following system of the ordinary differential equations
\cite{Arnold}
\ba\label{dina} \dot x_{n}=v_n(x), \ 1\leq n\leq N,
\ea
$\dot x_{n}$ stands for the total
derivative with respect to the parameter t.

When the number of the degrees of freedom is even,
and \be \label{Hams} v_n(x)=\varepsilon_{nm}\frac{\partial
H_0}{\partial x_m}, \ 1\leq n,m\leq 2M, \ee the system (\ref{dina})
is Hamiltonian one and can be put in the form \be \dot
x_{n}=\{x_n,H_0\}_0, \ee where the Poisson bracket is defined as
\be \{A,B\}_0=\varepsilon_{nm}\frac{\partial A}{\partial
x_n}\frac{\partial B} {\partial
x_m}=A\frac{\stackrel{\leftarrow}\partial }{\partial x_n}
\varepsilon_{nm}\frac{\stackrel{\rightarrow}\partial }{\partial
x_m}B, \ee and
summation rule under repeated indices has been used.

Let us consider the following Lagrangian
\be\label{LBD}
L=(\dot
x_{n}-v_n(x))\psi_n
\ee
and the corresponding equations of motion
\ba \label{din1} &&\dot x_{n}=v_n(x),
\dot
\psi_n=-\frac{\partial v_m}{\partial x_n}\psi_m. \ea The system
(\ref{din1}) extends the general system (\ref{dina}) by linear
equation for the variables $\psi$. The extended system can be put
in the Hamiltonian form
\cite{MV}
\ba \label{din1H}
&&\dot x_{n}=\{x_n,H_1\}_1,
\dot \psi_{n}=\{\psi_n,H_1\}_1,
\ea where first level (order) Hamiltonian is \be \label{H_1}
H_1=v_n(x)\psi_n \ee and (first level) bracket is defined as
\be\label{bb} \{A,B\}_1=A( \frac{\stackrel{\leftarrow}\partial
}{\partial x_n} \frac{\stackrel{\rightarrow}\partial }{\partial
\psi_n}- \frac{\stackrel{\leftarrow}\partial }{\partial \psi_n}
\frac{\stackrel{\rightarrow}\partial }{\partial x_n})B. \ee
Note
that when the Grassmann grading \cite{Berezin} of the conjugated
variables $x_n$ and $\psi_n$
are different, the bracket (\ref{bb}) is known as Buttin bracket\cite{Buttin}.\\

In the Faddeev-Jackiw formalism \cite{FJ} for the Hamiltonian
treatment of systems defined by first-order Lagrangians, i.e. by a Lagrangian
of the form
\ba
L=f_n(x)\dot{x}_n-H(x),
\ea
motion equations
\ba
f_{mn}\dot{x}_n=\frac{\partial H}{\partial x_m},
\ea
for the regular structure function $f_{mn},$ can be put in the explicit hamiltonian (Poisson; Dirac) form
 \ba
 \dot{x}_n=f^{-1}_{nm}\frac{\partial H}{\partial x_m}=\{x_n,x_m\}\frac{\partial H}{\partial x_m}=\{x_n,H\},
 \ea
 where the fundamental Poisson (Dirac) bracket is
 \ba
 \{x_n,x_m\}=f^{-1}_{nm}, \ f_{mn}=\partial_mf_n-\partial_nf_m.
 \ea
 The system (\ref{din1}) is an important example of the first order regular hamiltonian systems. Indeed, in the new variables,
\ba
y^1_n=x_n, y^2_n=\psi_n,
\ea
lagrangian (\ref{LBD}) takes the following first order form
\ba
&&L=(\dot
x_{n}-v_n(x))\psi_n \Rightarrow \frac{1}{2}(\dot
x_{n}\psi_n-\dot{\psi}_nx_n)-v_n(x)\psi_n=\frac{1}{2}y^a_n\varepsilon^{ab}\dot{y}_n^b-H(y)
\cr&&
=f_n^a(y)\dot{y}_n^a-H(y), f_n^a=\frac{1}{2}y^b_n\varepsilon^{ba}, H=v_n(y^1)y^2_n,
\cr &&
f^{ab}_{nm}=\frac{\partial f^b_m}{\partial y^a_n}-\frac{\partial f^a_n}{\partial y^b_m}=\varepsilon^{ab}\delta_{nm};
\ea
corresponding motion equations and the fundamental Poisson bracket  are
\ba
\dot{y}_n^a=\varepsilon_{ab}\delta_{nm}\frac{\partial H}{\partial y^b_m}=\{y_n^a,H\}, \{y^a_n,y^b_m\}=\varepsilon_{ab}\delta_{nm}.
\ea
 To the canonical quantization of this system corresponds
 \ba
 [\hat{y}^a_n,\hat{y}^b_m]=i\hbar\varepsilon_{ab}\delta_{nm},\  \hat{y}^1_n=y^1_n, \ \hat{y}^2_n=-i\hbar\frac{\partial}{\partial y^1_n}
 \ea
 In this quantum theory, classical part, motion equations for $y^1_n,$ remain classical.
 \subsection{Modified Bochner-Killing-Yano (MBKY) structures}

Now we return to our extended system (\ref{din1}) and formulate
conditions for the integrals of motion $H(x,\psi )$ \be\label{IH}
H=H_0(x)+H_1+...+H_N, \ee where \be\label{IHH}
H_n=A_{k_1k_2...k_n}(x)\psi_{k_1}\psi_{k_2}...\psi_{k_N}, \  1\leq
n\leq N, \ee we are assuming Grassmann valued $\psi_n$ and the
tensor $A_{k_1k_2...k_n}$ is skew-symmetric. For integrals
(\ref{IH}) we have \be \dot H=\{ \sum_{n=0}^{N}H_n,H_1\}
=\sum_{n=0}^{N}\{ H_n,H_1\} =\sum_{n=0}^{N} \dot H_n=0. \ee
Now we see, that each term in the sum (\ref{IH}) must be conserved separately.\\
In particular for Hamiltonian systems (\ref{Hams}), zeroth, $H_0$
and first level $H_1$, (\ref{H_1}), Hamiltonians are integrals of
motion. For $n=0$
\be \dot H_0=H_{0,k}v_k=0, \ee

for $1\leq n\leq N$ we have \ba\label{IHn} &&\dot H_n=\dot
A_{k_1k_2...k_n}\psi_{k_1}\psi_{k_2}...\psi_{k_N}+
 A_{k_1k_2...k_n}\dot \psi_{k_1}\psi_{k_2}...\psi_{k_N}+...
+ A_{k_1k_2...k_n}\psi_{k_1}\psi_{k_2}...\dot \psi_{k_N}\cr
&&=(A_{k_1k_2...k_n,k}v_k-A_{kk_2...k_n}v_{k_1,k}
 -...-A_{k_1...k_{n-1}k}v_{k_n,k})\psi_{k_1}\psi_{k_2}...\psi_{k_N}=0,
\ea and there is one-to-one correspondence between the existence
of the integrals (\ref{IHH}) and the existence of the nontrivial
solutions of the following equations \ba\label{MBKY}
\frac{D}{Dt}A_{k_1k_2...k_n}=
A_{k_1k_2...k_n,k}v_k-A_{kk_2...k_n}v_{k_1,k}
- ...-A_{k_1...k_{n-1}k}v_{k_n,k}
=0. \ea

For $n=1$ the system (\ref{MBKY}) gives \ba\label{MBKY1}
&&A_{k_1,k}v_k-A_{k}v_{k_1,k}=0 \ea and this equation has at list
one solution, $A_k=v_k.$ If we have two (or more) independent
first order integrals \ba H_1^{(1)}=A_k^{1}\Psi_k;\
H_1^{(2)}=A_k^{2}\Psi_k,... \ea we can construct corresponding
(reducible) second (or higher)order MBKY tensor(s) \ba
&&H_2=H_1^{(1)}H_1^{(2)}={A_{k}^{1}A_l^{2}}\Psi_k\Psi_l=A_{kl}\Psi_k\Psi_l;\cr
&&H_M=H_1^{(1)}...H_M^{(M)}=A_{k_1...k_M}\Psi_{k_1}...\Psi_{k_M},
\cr &&A_{k_1...k_M}=\{A_{k_1}^{(1)}...A_{k_M}^{(M)}\},\ 2\leq
M\leq N \ea
where under the bracket
operation, $\{B_{k_1,...,k_N}\}=\{B\}$ we understand complete
anti-symmetrization.
 The system (\ref{MBKY}) defines a Generalization of the
Bochner-Killing-Yano structures
of the
geodesic motion of the point particle, for the case of the general
(\ref{dina}) (and extended (\ref{din1})) dynamical systems. Having
$A_M, 2\leq M\leq N$ independent MBKY structures, we can construct
corresponding second order Killing tensors and Nambu-Poisson
dynamics. In the superintegrable case, we have maximal number of
the motion integrals, N-1.

The structures defined by the system (\ref{MBKY}) we
call  the Modified Bochner-Killing-Yano structures or MBKY structures for short, \cite{Makhaldiani1999}.

 \subsection{Point vortex dynamics (PVD)}
 PVD can dy defined (see e.g.
 \cite{vort,vort1}
 ) as the following first order system
 \ba\label{Nvort}
 \dot{z}_n=i\sum_{m\neq n}^N\frac{\gamma_m}{z^*_n-z^*_m},\  z_n=x_n+iy_n,\ 1\leq n\leq N.
 \ea
Corresponding first order lagrangian, hamiltonian, momenta, Poisson brackets and commutators are
\ba
&&L=\sum_n \frac{i}{2}\gamma_n(z_n\dot{z}_n^*-\dot{z}_nz_n^*)-\sum_{n\neq m}\gamma_n\gamma_m ln|z_n-z_m|\cr
&&H=\sum_{n\neq m}\gamma_n\gamma_m\ln |z_n-z_m|\cr
&&=\frac{1}{2}\sum_{n\neq m}\gamma_n\gamma_m(\ln (z_n-z_m)+\ln(p_n-p_m)),\cr
&& p_n=\frac{\partial L}{\partial \dot{z}_n}=-\frac{i}{2}\gamma_nz_n^*,\ p_n^*=\frac{\partial L}{\partial \dot{z}_n^*}=\frac{i}{2}\gamma_nz_n,\cr
&&\{p_n,z_m\}=\delta_{nm},\ \{p^*_n,z^*_m\}=\delta_{nm},\ \{x_n,y_m\}=\delta_{nm},\cr
&&[p_n,z_m]=-i\hbar\delta_{nm}\Rightarrow[x_n,y_m]=-i\frac{\hbar}{\gamma_n}\delta_{nm}
\ea
So, quantum vortex dynamics corresponds to the noncommutative space. It is natural to assume that vortex parameters are
quantized as
\ba
\gamma_n=\frac{\hbar}{a^2}n,\ n=\pm1,\pm2,...
\ea
and $a$ is a characteristic (fundamental) length.

\section{Nambu dynamics}
In the canonical formulation,
the equations of motion
of a physical system are defined via a Poisson
bracket and a Hamiltonian, \cite{Arnold}. In Nambu's formulation, the Poisson bracket is replaced by the Nambu
bracket with $n+1, n\geq1,$ slots. For $n=1,$ we have the canonical formalism with one Hamiltonian. For $n\geq2,$ we have Nambu-Poisson formalism, with $n$ Hamiltonians, \cite{Nambu},
\cite{Whittaker}.


\subsection{System of three vortexes}

The system of $N$ vortexes (\ref{Nvort})
for $N=3,$ and
\begin{eqnarray}
u_1=ln|z_2-z_3|^2,
u_2=ln|z_3-z_1|^2,
u_3=ln|z_1-z_2|^2
\end{eqnarray}
reduce to the following system
\begin{eqnarray}  \label{vor}
\dot u_{1}=\gamma_{1}(e^{u_{2}}-e^{u_{3}}),
{\dot u_{2}}
=\gamma_{2}(e^{u_{3}}-e^{u_{1}}),
{\dot u_{3}}%
=\gamma_{3}(e^{u_{1}}-e^{u_{2}}),
\end{eqnarray}


The system (\ref{vor}) has two integrals of motion
\begin{eqnarray}  \label{hami}
H_{1}=\sum_{i=1}^{3}\frac{e^{u_i}}{\gamma_i},
H_{2}=\sum_{i=1}^{3}\frac{u_i}{\gamma_i}  \nonumber
\end{eqnarray}
\noindent and can be presented in the Nambu--Poisson form,
\cite{M1}
\begin{eqnarray}  \label{nambu1}
\dot u_{i}=\omega_{ijk}\frac{\partial H_1}{\partial u_j} \frac{\partial H_2%
}{\partial u_k}
=\{x_{i},H_1,H_2\}=\omega_{ijk}\frac{e^{u_j}}{\gamma_{j}} \frac{1}{%
\gamma_{k}},  \nonumber
\end{eqnarray}
where
\begin{eqnarray}
\omega_{ijk}=\epsilon_{ijk}\rho,
\rho=\gamma_{1}\gamma_{2}\gamma_{3}  \nonumber
\end{eqnarray}
and the Nambu--Poisson bracket of the functions $A,B,C$ on the
three-dimensional phase space is
\begin{eqnarray}  \label{nambu2}
\{ A,B,C\}=\omega_{ijk}\frac{\partial A}{\partial u_i}\frac{\partial B} {%
\partial u_j}\frac{\partial C}{\partial u_k}.
\end{eqnarray}



This system is superintegrable: for $N=3$ degrees of
freedom, we have maximal number of the integrals of motion
$N-1=2.$
\subsection{Extended quantum mechanics
}
As an example of the infinite dimensional Nambu-Poisson dynamics,
let me conside the following extension of  Schr$\ddot o$dinger
quantum mechanics \cite{Makhaldiani2000}\ba\label{ES}
&&iV_t=\Delta V-\frac{V^2}{2},\label{V} \\
&&i\psi_t=-\Delta \psi+V\psi.
\ea An interesting
solution to the equation for the potential (\ref{V}) is
\ba\label{MQM}
V=\frac{4(4-d)}{r^2},
\ea
where $d$ is the dimension of the spase.
In the case of $d=1,$ we have the potential of conformal quantum
mechanics.

The variational formulation of the extended quantum theory,
is given by the following Lagrangian
\ba \label{eLSh} L=(iV_t-\Delta V+\frac{1}{2}V^2)\psi. \ea
The
momentum variables are
\ba &&P_v=\frac{\partial L}{\partial
V_t}=i\psi,
P_{\psi}=0. \ea As  Hamiltonians of the
Nambu-theoretic formulation, we take the following integrals of
motion \ba &&H_1=\int d^dx(\Delta V-\frac{1}{2}V^2)\psi,\cr
&&H_2=\int d^dx(P_v-i\psi),\cr &&H_3=\int d^dx P_{\psi}. \ea We
invent unifying vector notation,
$\phi=(\phi_1,\phi_2,\phi_3,\phi_4)= (\psi,P_{\psi},V,P_v).$ Then
it may be verified that the equations of the extended quantum
theory can be put in the following Nambu-theoretic form
\ba\label{NSH} &&\phi_t(x)=\{\phi(x),H_1,H_2,H_3\},
\ea

where the bracket is defined as \ba
&&\{A_1,A_2,A_3,A_4\}
=i\varepsilon_{ijkl} \int \frac{\delta
A_1}{\delta \phi_i(y)} \frac{\delta A_2}{\delta
\phi_j(y)}\frac{\delta A_3}{\delta \phi_k(y)} \frac{\delta
A_4}{\delta \phi_l(y)}dy\cr
&&=i\int \frac{\delta (A_1, A_2, A_3,
A_4)}{\delta (\phi_1(y),\phi_2(y),\phi_3(y),\phi_4(y))}dy
=idet(\frac{\delta A_k}{\delta \phi_l}). \ea

\subsection{$M$ theory}
The basic building blocks of M theory are membranes
and $M5-$branes. Membranes are fundamental objects carrying electric charges with respect to
the 3-form $C$-field, and $M5$-branes are magnetic solitons.
The Nambu-Poisson
3-algebras
appear as gauge symmetries of
superconformal Chern-Simons nonabelian
theories in 2 + 1 dimensions with the maximum allowed number of $N = 8$
linear supersymmetries.



The Bagger and Lambert $\cite{BaggerLambert}$  and,  Gustavsson $\cite{Gustavsson}$ (BLG) model
is based on a
3-algebra,
\ba
[T^a,T^b,T^c]=f^{abc}_dT^d
\ea
where $T^a$, are generators and $f_{abcd}$ is a fully anti-symmetric tensor. Given this algebra, a maximally supersymmetric
Chern-Simons lagrangian is:
\ba
&&L=L_{CS}+L_{matter}, L_{CS}=\frac{1}{2}\varepsilon^{\mu\nu\lambda}(f_{abcd}A^{ab}_\mu\partial_\nu A^{cd}_\lambda+\frac{2}{3}f_{cdag}f_{efb}^gA^{ab}_\mu A^{cd}_\nu A^{ef}_\lambda),\cr
&&L_{matter}=
\frac{1}{2}B_\mu ^{Ia}B^{\mu I}_{a}-B_\mu ^{Ia}D^\mu X^I_a+
+\frac{i}{2}\bar{\psi}^a\Gamma^\mu D_\mu \psi_a
+\frac{i}{4}\bar{\psi}^b\Gamma_{IJ}x_c^Ix_d^J\psi_af^{abcd}\cr
&&-\frac{1}{12}tr([X^I,X^J,X^K][X^I,X^J,X^K]),\ I=1,2,...,8,
\ea
where $A_\mu^{ab}$ is gauge boson, $\psi^a$ and $X^I=X^I_aT^a$ matter fields. If
$a = 1,2,3,4,$ then we can obtain an $SO(4)$ gauge symmetry by choosing $f_{abcd} = f\varepsilon_{abcd}, f$ being
a constant. It turns out to be the only
case that gives a gauge theory with manifest unitarity and $N = 8$ supersymmetry.

The action has the first order form so we can use the formalism of the first section.
The motion equations for the gauge fields
\ba
f^{nm}_{abcd}\dot{A}_m^{cd}(t,x)=\frac{\delta H}{\delta A_n^{ab}(t,x)}, f^{nm}_{abcd}=\varepsilon^{nm}f_{abcd}
\ea
take canonical form
\ba
&&\dot{A}_n^{ab}=f^{abcd}_{nm}\frac{\delta H}{\delta A_m^{cd}}=\{A_n^{ab},A_m^{cd}\}\frac{\delta H}{\delta A_m^{cd}}=\{A_n^{ab},H\},\cr
&&\{A_n^{ab}(t,x),A_m^{cd}(t,y)\}=\varepsilon_{nm}f^{abcd}\delta^{(2)}(x-y)
\ea



\section{Discrete dynamical systems}\label{sec2}
Computers are physical devices and their behavior is determined by physical laws.
The Quantum Computations \cite{QuanputRef1,QuanputRef2}, Quantum Computing, Quanputing \cite{Quanputers}, is
a new interdisciplinary field of research, which benefits from the contributions of physicists, computer scientists, mathematicians, chemists and engineers.

Contemporary digital computer and its logical elements
can be considered as a spatial type of discrete dynamical systems
\cite{QP1}
\begin{eqnarray}  \label{ds}
S_n(k+1)=\Phi_n (S(k)),
\end{eqnarray}
where \ba S_n(k), \ \ 1\leq n\leq N(k), \ea
 is the state vector of
the system at the discrete time step $k$. Vector $S$ may describe
the
state 
 and $\Phi$ transition
rule of some Cellular Automata \cite{CA}.The systems of the type
(\ref{ds}) appears in applied mathematics as an explicit finite
difference scheme approximation of the equations of the physics
\cite{Samarski}.

{\sl \bf Definition:} {\it We assume that the system (\ref{ds}) is
time-reversible if we can define the reverse dynamical system}
\begin{eqnarray}  \label{ids}
S_n(k)=\Phi_n^{-1} (S(k+1)).
\end{eqnarray}
In this case the following matrix
\begin{eqnarray}
M_{nm}=\frac{\partial \Phi_n(S(k))}{\partial S_m(k)},
\end{eqnarray}
is regular, i.e. has an inverse.
 If the matrix is not regular, this is the case, for example, when
$N(k+1)\neq N(k),$
 we have an irreversible dynamical system
(usual digital computers and/or corresponding irreversible gates).

Let us consider an extension of the dynamical system (\ref{ds})
given by the following action function
\begin{eqnarray}\label{A}
A=\sum_{kn}^{}l_n(k)(S_n(k+1)-\Phi_n(S(k)))
\end{eqnarray}
and corresponding motion equations
\begin{eqnarray}\label{sl}
&&S_n(k+1)=\Phi_n (S(k))=\frac{\partial H}{\partial l_n(k)},\cr
&&l_n(k-1)=l_m(k)\frac{\partial \Phi_m(S(k))}{%
\partial S_n(k)} =l_m(k)M_{mn}(S(k))=\frac{\partial H}{\partial S_n(k)},
\end{eqnarray}
where
\ba H=\sum_{kn}l_n(k)\Phi_n(S(k)),
\ea
is discrete
Hamiltonian.
 In the regular case, we put the system (\ref{sl}) in an explicit form
\begin{eqnarray}  \label{eds}
&&S_n(k+1)=\Phi_n (S(k)),\cr &&l_n(k+1)=l_m(k)M^{-1}_{mn}(S(k+1)).
\end{eqnarray}

From this system it is obvious that, when the initial value
$l_n(k_0)$ is given, the evolution of the vector $l(k)$ is defined by
evolution of the state vector $S(k).$ The equation of motion for
$l_n(k)$ - Elenka is linear and has an important property that a
linear superpositions of the solutions are also solutions.

 {\bf Statement:} {\it Any time-reversible dynamical system (e.g. a
time-reversible computer) can be extended by corresponding linear
dynamical system (quantum - like processor) which is controlled by
the dynamical system and
has a huge computational power,}
\cite{QP1,QP2,Quanputers,Quanputers2}.



\subsection{ (de)Coherence criterion.}
For motion equations (\ref{sl}) in the continual
approximation, we have
\begin{eqnarray}
&&S_n(k+1)=x_n(t_k+\tau)=x_n(t_k)+\dot{x}_n(t_k)\tau+O(\tau^2)
,\cr &&\dot{x}_n(t_k)=v_n(x(t_k))+O(\tau),\ \ t_k=k\tau, \cr
&&v_n(x(t_k))=(\Phi_n (x(t_k))-x_n(t_k))/\tau ;\cr
&&M_{mn}(x(t_k))=\delta_{mn}+\tau\frac{\partial
v_m(x(t_k))}{\partial x_n(t_k)}.
\end{eqnarray}
{\bf (de)Coherence criterion:} {\it the system is reversible, the
linear (quantum, coherent, soul) subsystem exists, when the matrix
$M$ is regular,} \ba detM=1+\tau \sum_n \frac{\partial
v_n}{\partial x_n} +O(\tau^2)\neq 0. \ea For the Nambu - Poisson
dynamical systems (see e.g. \cite{NPD}) \ba
&&v_n(x)=\varepsilon_{nm_1m_2...m_p}\frac{\partial H_1}{\partial
x_{m_1}}\frac{\partial H_2}{\partial x_{m_2}}...\frac{\partial
H_p}{\partial x_{m_p}},\ \ p=1,2,3, ...,N-1,\cr && \sum_n
\frac{\partial v_n}{\partial x_n}\equiv divv=0.\ea


\end{document}